\documentclass[epj,nopacs]{svjour}
\pdfoutput=1
\usepackage{graphicx,hyperref,wrapfig,cite,color,subfigure,amssymb,amsmath,epsfig}
\usepackage[utf8]{inputenc}
\usepackage[english]{babel}
\hypersetup{
     colorlinks=true,        
     linkcolor=red,          
     citecolor=green,        
     filecolor=magenta,      
     urlcolor=blue           
}
\newcommand{\newc}{\newcommand}
\def\u#1{\verb!#1!\endgroup}
\newc{\HW}{\mbox{\textsf{HERWIG}}}
\newc{\Hw}{\mbox{\textsf{Herwig}}}
\newc{\KrkNLO}{\textsf{KrkNLO}}
\newc{\OL}{\textsf{OpenLoops}}
\newc{\Collier}{\textsf{Collier}}
\newc{\GoSam}{\textsf{GoSam}}
\newc{\TAUOLA}{\textsf{TAUOLA}}
\newc{\ThePEG}{\textsf{ThePEG}}
\newc{\boost}{\textsf{BOOST}}
\newc{\HepMC}{\textsf{HepMC}}
\newc{\Rivet}{\textsf{Rivet}}
\newc{\lhapdf}{\textsf{LHAPDF}}
\newc{\HWPP}{\mbox{\textsf{Herwig++}}}
\newc{\evt}{\textsf{EvtGen}}
\newc{\fortran}{\textsf{FORTRAN}}
\newc{\decayer}{\textsf{Decayer}}
\newc{\matchbox}{\textsf{Matchbox}}

\newc{\HWPPClass}[1]{\href{https://herwig.hepforge.org/doxygen/classHerwig_1_1#1.html}{\textsf{#1}}}
\newc{\ThePEGClass}[1]{\href{https://thepeg.hepforge.org/doxygen/classThePEG_1_1#1.html}{\textsf{#1}}}
\newc{\HWPPParameter}[2]{\href{https://herwig.hepforge.org/doxygen/#1Interfaces.html\##2}{{\bf #2}}}
\newc{\ThePEGParameter}[2]{\href{https://thepeg.hepforge.org/doxygen/#1Interfaces.html\##2}{{\bf #2}}}
\newc{\HWPPParameterValue}[3]{\href{https://herwig.hepforge.org/doxygen/#1Interfaces.html\##2}{{\bf [#2=#3]}}}
\newc{\HWPPParameterValueB}[3]{\href{https://herwig.hepforge.org/doxygen/#1Interfaces.html\##2}{{\bf [#3]}}}
\newc{\ThePEGParameterValue}[3]{\href{https://thepeg.hepforge.org/doxygen/#1Interfaces.html\##2}{{\bf [#2=#3]}}}

\preprint{
MAN/HEP/2019/011\\
CERN-TH-2019-213\\ 
IFJPAN-IV-2019-18\\
HERWIG-2019-02\\
UWTHPH-19-36\\
KA-TP-24-2019\\
LU-TP 19-57\\
MCnet-19-28\\
IPPP/19/91
}

\title{Herwig 7.2 Release Note}

\author{
  Johannes~Bellm\inst{1}\and
  Gavin~Bewick\inst{2}\and
  Silvia~Ferrario~Ravasio\inst{2}\and
  Stefan~Gieseke\inst{3}\and
  David~Grellscheid\inst{4}\and
  Patrick~Kirchgae\ss{}er\inst{3}\and
  Mohammad~R.~Masouminia\inst{2}\and
  Graeme~Nail\inst{5}\and
  Andreas~Papaefstathiou\inst{5}\and
  Simon~Pl\"atzer\inst{6}\and
  Michael~Rauch\inst{3}\and
  Christian~Reuschle\inst{1}\and
  Peter~Richardson\inst{2,7}\and
  Michael~H.~Seymour\inst{8}\and
  Andrzej~Si\'odmok\inst{9}\and
  Stephen~Webster\inst{2}
}
\institute{
  Department of Astronomy and Theoretical Physics, Lund University,\and
  IPPP, Department of Physics, Durham University,\and
  Institute for Theoretical Physics, Karlsruhe Institute of Technology,\and
  Department of Informatics, University of Bergen,\and
  Higgs Centre for Theoretical Physics, University of Edinburgh,\and
  Particle Physics, Faculty of Physics, University of Vienna,\and
  CERN, PH-TH, Geneva,\and
  Particle Physics Group, Department of Physics and Astronomy, University of
  Manchester,\and
  The Henryk Niewodniczański Institute of Nuclear Physics in Cracow, Polish
  Academy of Sciences.
}

\date{December 12, 2019}

\abstract{A new release of the Monte Carlo event generator Herwig (version
  7.2) is now available. This version introduces a number of improvements,
  notably: improvements to the simulation of multiple-parton interactions, including diffractive processes; a new model for baryonic colour
  reconnection; spin correlations in both the dipole and angular-ordered parton showers; improvements to strangeness production;
  an improved choice of evolution variable in the angular-ordered parton shower; support for
  generic Lorentz structures in BSM models.}

\begin{document}\sloppy

\maketitle

\section{Introduction}

\Hw\ is a multi purpose particle physics event generator.
The current version series, \Hw7\cite{Bellm:2015jjp}, is based on a major development
of the \HWPP
\cite{Bahr:2008pv,Bahr:2008tx,Bahr:2008tf,Gieseke:2011na,Arnold:2012fq,Bellm:2013lba} branch.
It fully supersedes the \HWPP\ 2.x and
\HW\ 6.x versions. Building on the technology and experience gained with the
higher-order improvements provided by \Hw\ 7.0\cite{Bellm:2015jjp} and 7.1\cite{Bellm:2017bvx}, a major follow-up release,
\Hw\ 7.2 is now available. The new version includes several improvements
to the soft components of the simulation,
amongst other changes and physics capabilities, which we will
highlight in this release note. Please refer to the \HWPP\ manual
\cite{Bahr:2008pv}, the \Hw\ 7.0 \cite{Bellm:2015jjp} as well as this release
note when using the new version of the program.
Studies or analyses that rely on a particular feature of the program should
also reference the paper(s) where the physics of that feature was first
described.
The authors are happy to provide guidance on which features are relevant
for a particular analysis.

\subsection{Availability}

The new version, as well as older versions of the \Hw\ event generator can be
downloaded from the website
\texttt{\href{https://herwig.hepforge.org/}{https://herwig.hepforge.org/}}.
We strongly recommend using the \texttt{bootstrap} script provided for the
convenient installation of \Hw\ and all of its dependencies, which can be
obtained from the same location. On the website, tutorials and FAQ sections are
provided to help with the usage of the program. Further enquiries should be
directed to \texttt{herwig@projects.hepforge.org}.  \Hw\ is released under the
GNU General Public License (GPL) version 3 and the MCnet guidelines for the
distribution and usage of event generator software in an academic setting, see
the source code archive or
\texttt{\href{http://www.montecarlonet.org/index.php?p=Publications/Guidelines}
  {http://www.montecarlonet.org/}}.

\subsection{Prerequisites and Further Details}

\Hw\ 7.2 is built on the same backbone and dependencies as its predecessors
\Hw\ 7.0 and 7.1, and uses the same method of build, installation and run
environment. No major changes should hence be required in comparison to a
working \Hw\ 7.1 installation. Some of the changes, though,
might require different compiler versions.
The tutorials at
\texttt{\href{https://herwig.hepforge.org/tutorials/}{https://herwig.hepforge.org/tutorials/}}
have been extended and adapted to the new version and serve as the primary
reference for physics setups and as a user manual until a comprehensive
replacement for the detailed manual \cite{Bahr:2008pv} is available.

\pagebreak
\section{Angular-Ordered Parton Shower}

A major restructuring of the angular-ordered parton shower has been
performed in order to simplify the code, remove unused levels of
abstraction and unused options. This is intended to improve the
maintainability of the code and make new developments easier.

\begin{figure}[tb]
\includegraphics[width=0.45\textwidth]{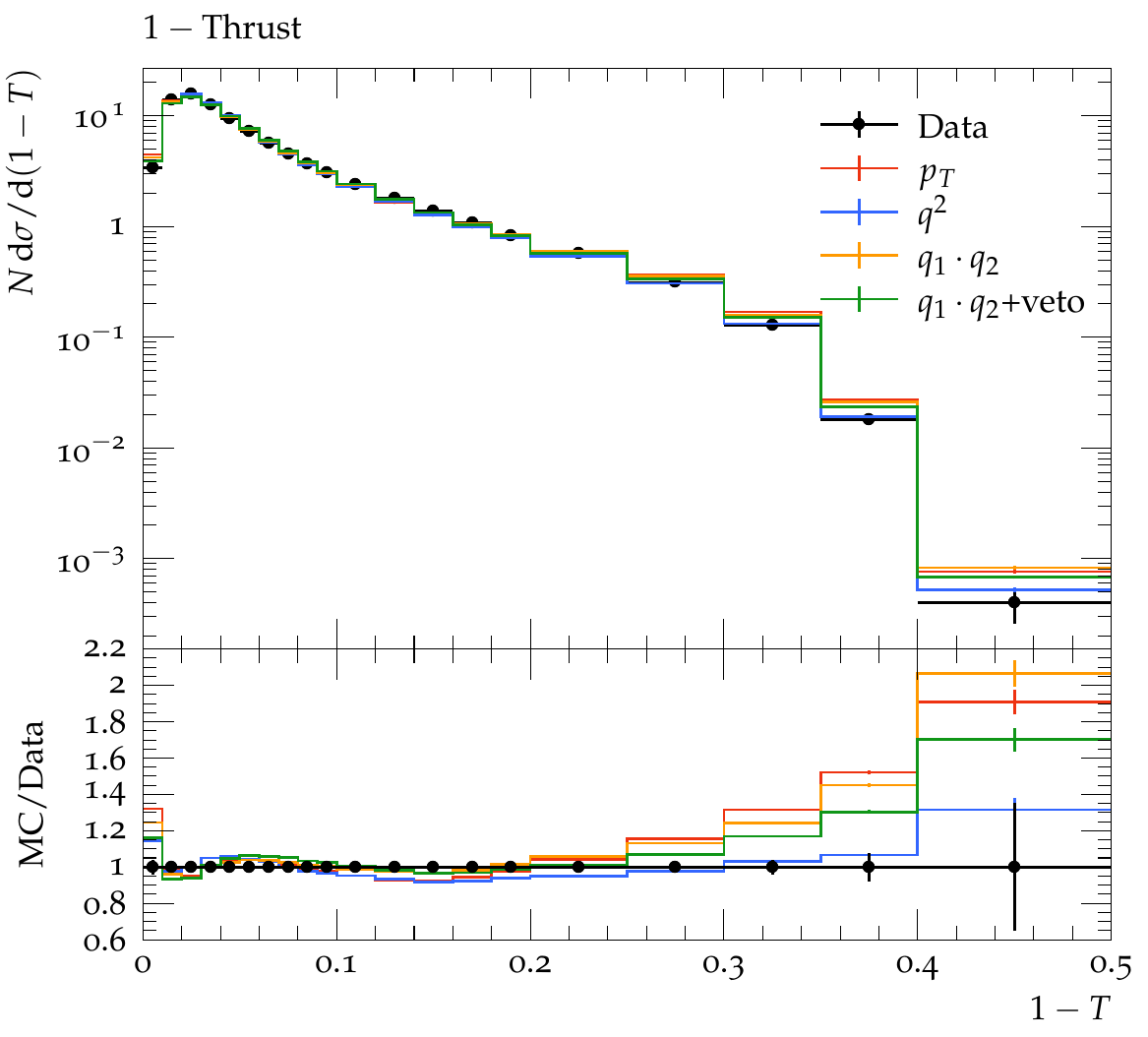}\\
\includegraphics[width=0.45\textwidth]{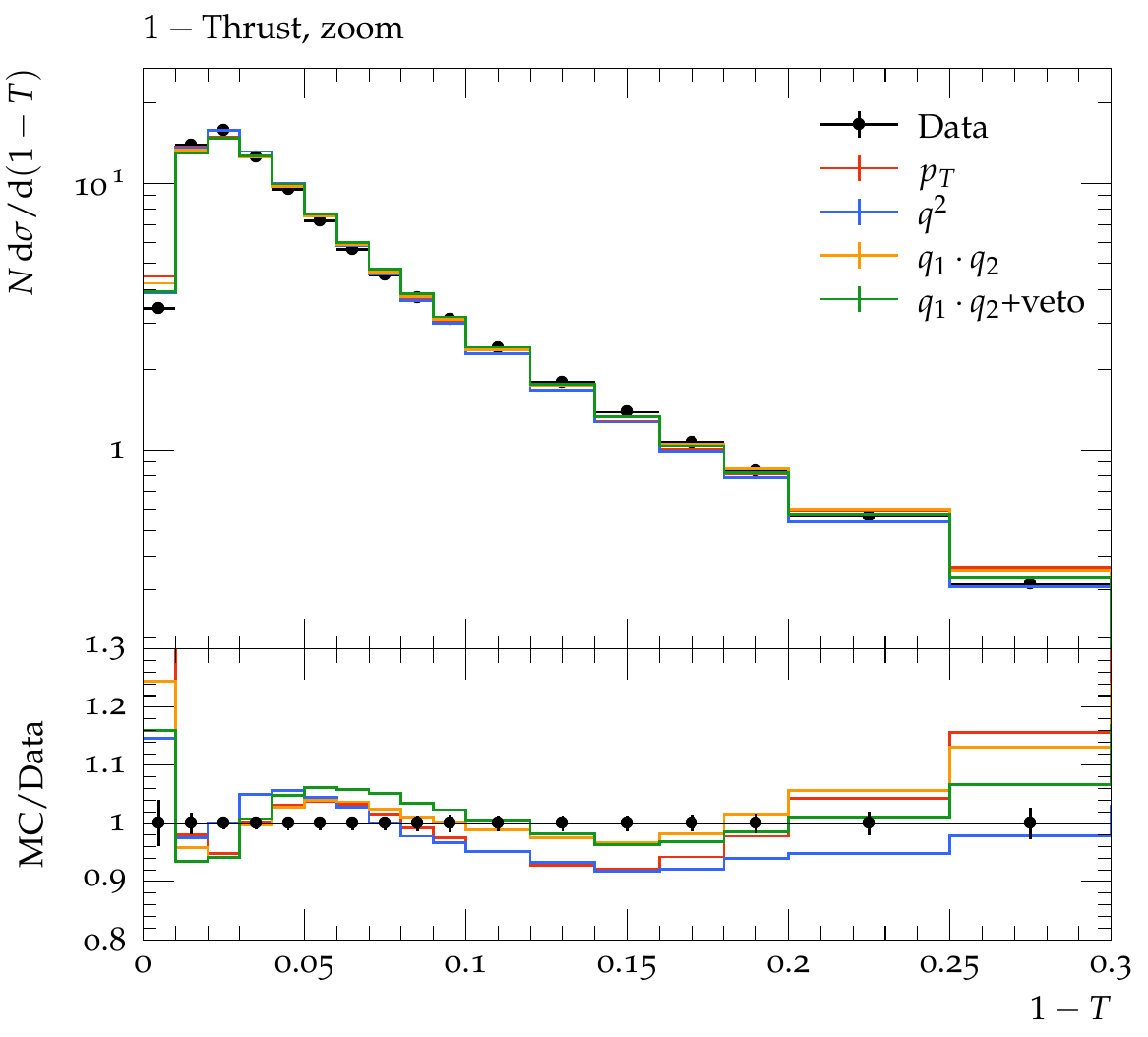}
\caption{The thrust at the Z-pole compared with data from the DELPHI~\cite{Abreu:1996na} experiment. In the right panel a zoom for small $1-T$ values is shown.}
\label{fig:thrust-DELPHI}
\end{figure}

In addition we have changed the default interpretation of the ordering
variable.  When a final-state splitting $i \to j,k$ is generated, we can define
the ordering scale in three different ways:
\begin{align}
\tilde{q}^2 &= \frac{q_i^2-m^2_i}{z(1-z)}; \label{eq:q2-preserving} \\
            &= \frac{p_T^2+(1-z)m^2_j+zm_k^2-z(1-z)m^2_i}{z^2(1-z)^2} \label{eq:pt-preserving}; \\
            &= \frac{2 q_j \cdot q_k +m_j^2+m_k^2 -m^2_i}{z(1-z)} \label{eq:dot-preserving};
\end{align}
where $z$ is the light-cone momentum fraction carried by the particle
$j$, $p_T$ is the transverse momentum of the splitting.  When multiple
emissions occur just one definition can be employed and this choice
will also determine which quantity is preserved.  We call this choice
the ``recoil scheme''. By default, the
scale is now expressed in terms of the dot-product of the emitted
particles, \emph{i.e.} Eq.~\eqref{eq:dot-preserving}, as discussed in
Ref.\,\cite{Bewick:2019rbu}.  We also include a veto on the masses of
final-state jets, as suggested in Ref.\,\cite{Bewick:2019rbu}, and we
adopt the tuned parameter obtained in Ref.\,\cite{Bewick:2019rbu}.
All of the choices for the interpretation of the evolution variable
and tunes from Ref.\,\cite{Bewick:2019rbu} are available using the
snippets
\begin{verbatim}
EvolutionScheme-*.in  Tune-*.in 
\end{verbatim}
where \texttt{*} can be \texttt{DotProduct-Veto}, \texttt{DotProduct},
\texttt{pT} or \texttt{Q2}.  This new recoil scheme, together with the
veto on the final-state jets, allows a better description of the
double-logarithmically enhanced region, without overpopulating the
tail of the distributions, as can be seen in
Fig.~\ref{fig:thrust-DELPHI} where the thrust distribution at the Z
pole is compared to LEP data. The $q^2$-preserving scheme~(blue)
yields a good description of the tail, while the
$p_T$-preserving~(red) one performs better in the $T \approx 1$
region, however the dot-product-preserving scheme, together with the
veto~(green), gives the best agreement with data over the whole range.

\section{Colour Matrix Element Corrections}

General colour matrix element corrections for the dipole shower as
presented in \cite{Platzer:2018pmd} and earlier outlined in
\cite{Platzer:2012np} are now available in the new release. The colour
matrix element corrections change the radiation pattern of the dipole
shower for subsequent emissions by including a correction factor
\begin{equation}
  w_{ij,k} = -\frac{{\rm Tr}[{\mathbf T}_{ij}\cdot {\mathbf T}_k\ {\mathbf M}_n]}{{\mathbf T}_{ij}^2 {\rm Tr}[{\mathbf M}_n]}
\end{equation}
along with each dipole splitting kernel $V_{ij,k}$, where ${\mathbf
  M}_n$ is the $n$-parton `colour density operator' initialized by
the amplitude and conjugate amplitude vectors at the level of the hard
process which is evolved to higher multiplicities using the
soft-collinear approximation. They can be enabled using the dipole
shower with any of the \texttt{Matchbox} generated processes and the
\begin{verbatim}
Matchbox/CMEC.in
\end{verbatim}
snippet.

\section{Spin Correlations}

\begin{figure}[tb]
\includegraphics[width=0.45\textwidth]{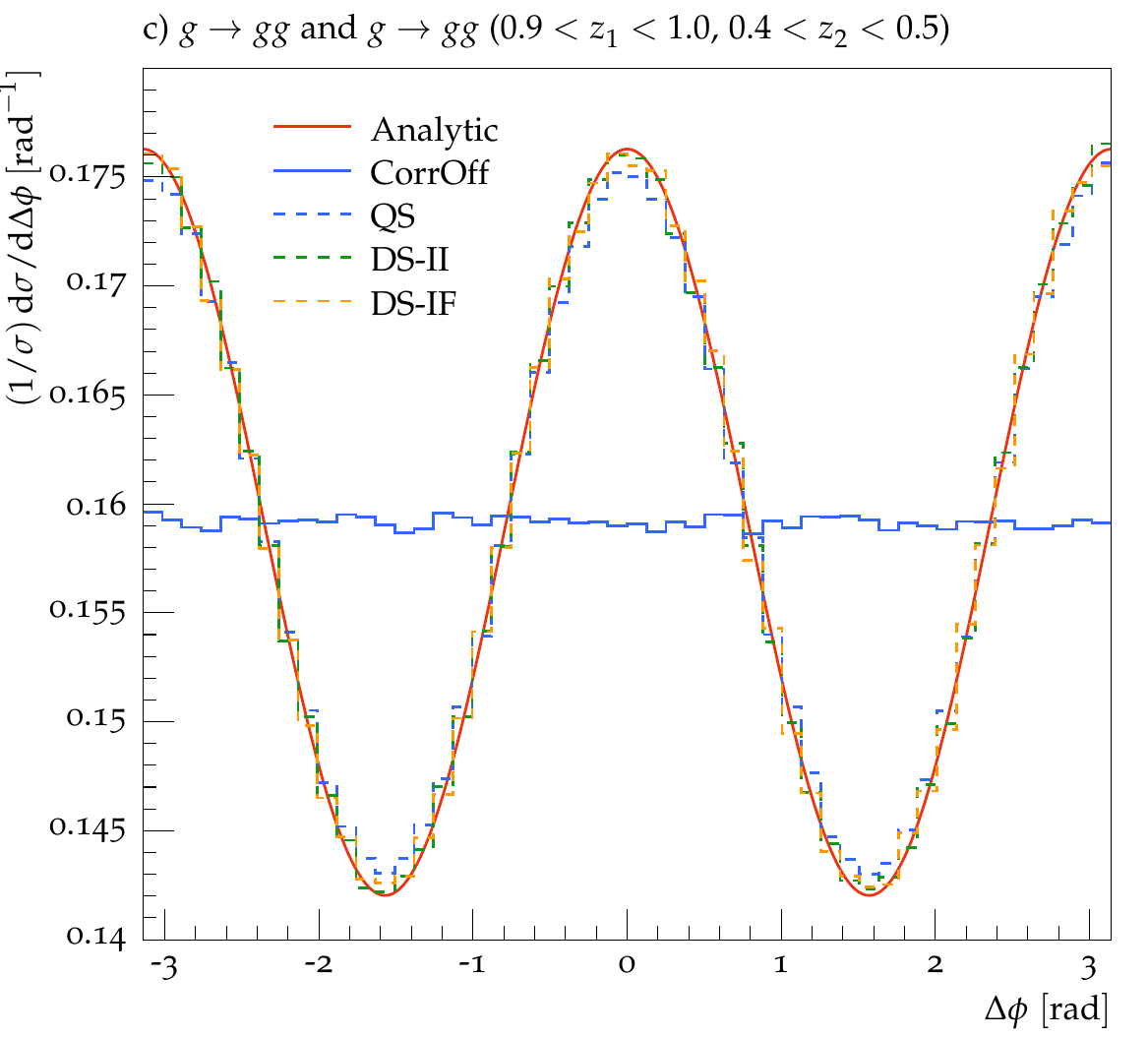}\\
\includegraphics[width=0.45\textwidth]{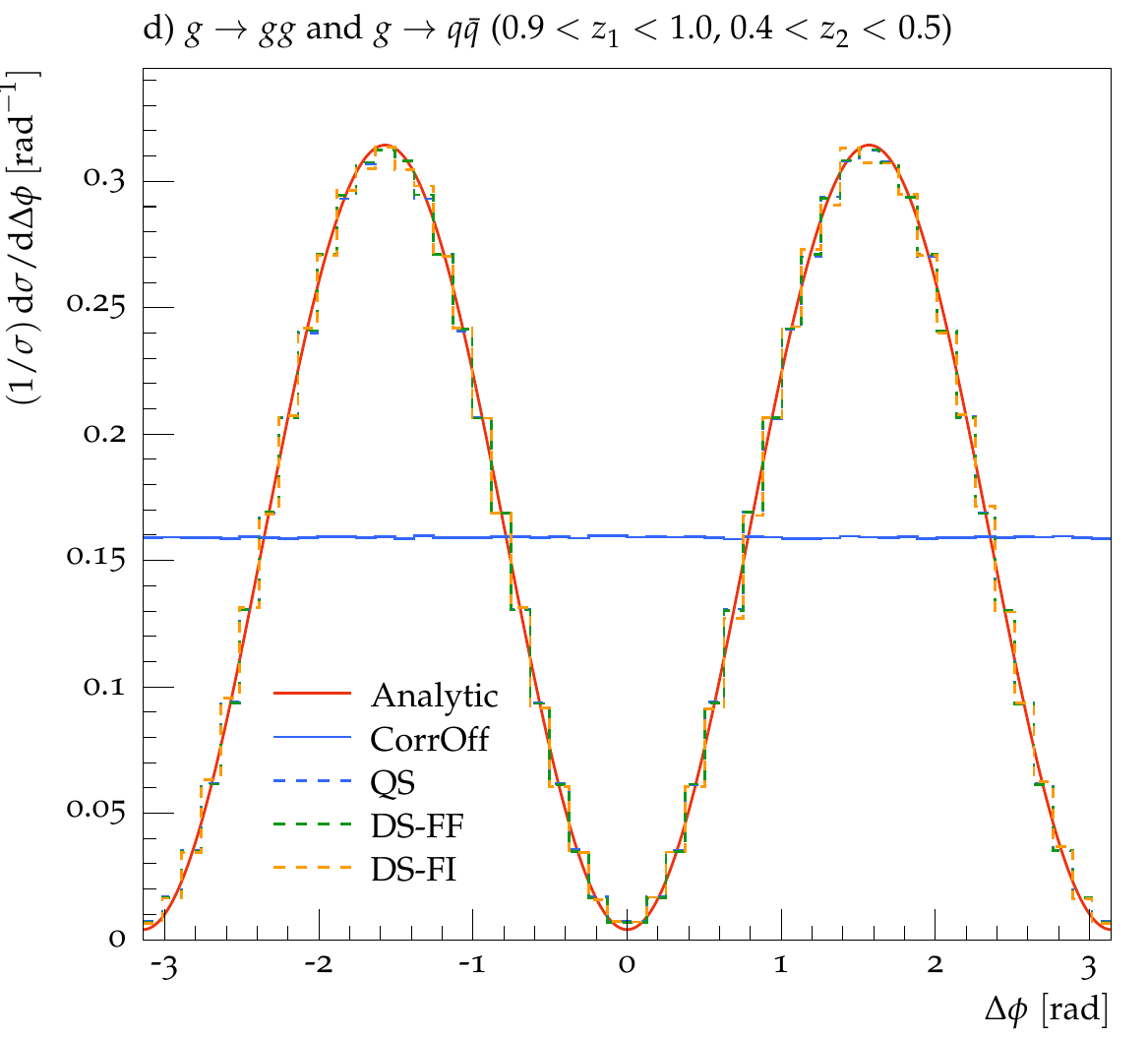}
\caption{Examples of the spin correlations in the parton shower for $g\to gg$ with subsequent $g\to gg$ and $g\to q\bar{q}$ branching. For
details, see Ref.~\cite{Richardson:2018pvo}.}
\label{fig:spin}
\end{figure}

\Hw7 has always included spin correlations between production and decay of particles,
and in both perturbative and non-perturbative decays. We have now completed the inclusion of
spin correlations in all stages of the event generation by incorporating the correlations
into both the angular-ordered and dipole parton showers. An example of these correlations is
shown in Fig.\,\ref{fig:spin} and this work is described in more detail in Ref.\,\cite{Richardson:2018pvo}.

\section{Perturbative Decays}

The classes implementing perturbative decays, in both the Standard Model and for BSM models, have been
restructured. This allows the several previous implementations of hard radiation corrections in these decays,
in both the POWHEG and matrix element correction schemes, to be combined and generalised.
This now allows us to apply POWHEG-style hard corrections to a much wider range of decays, in particular in BSM models, and also include hard QED radiation. This restructuring also allows these decays, and the POWHEG corrections, to be used with both parton shower modules.

\section{Baryonic Colour Reconnection}
\begin{figure}[t]
\includegraphics[width=0.48\textwidth]{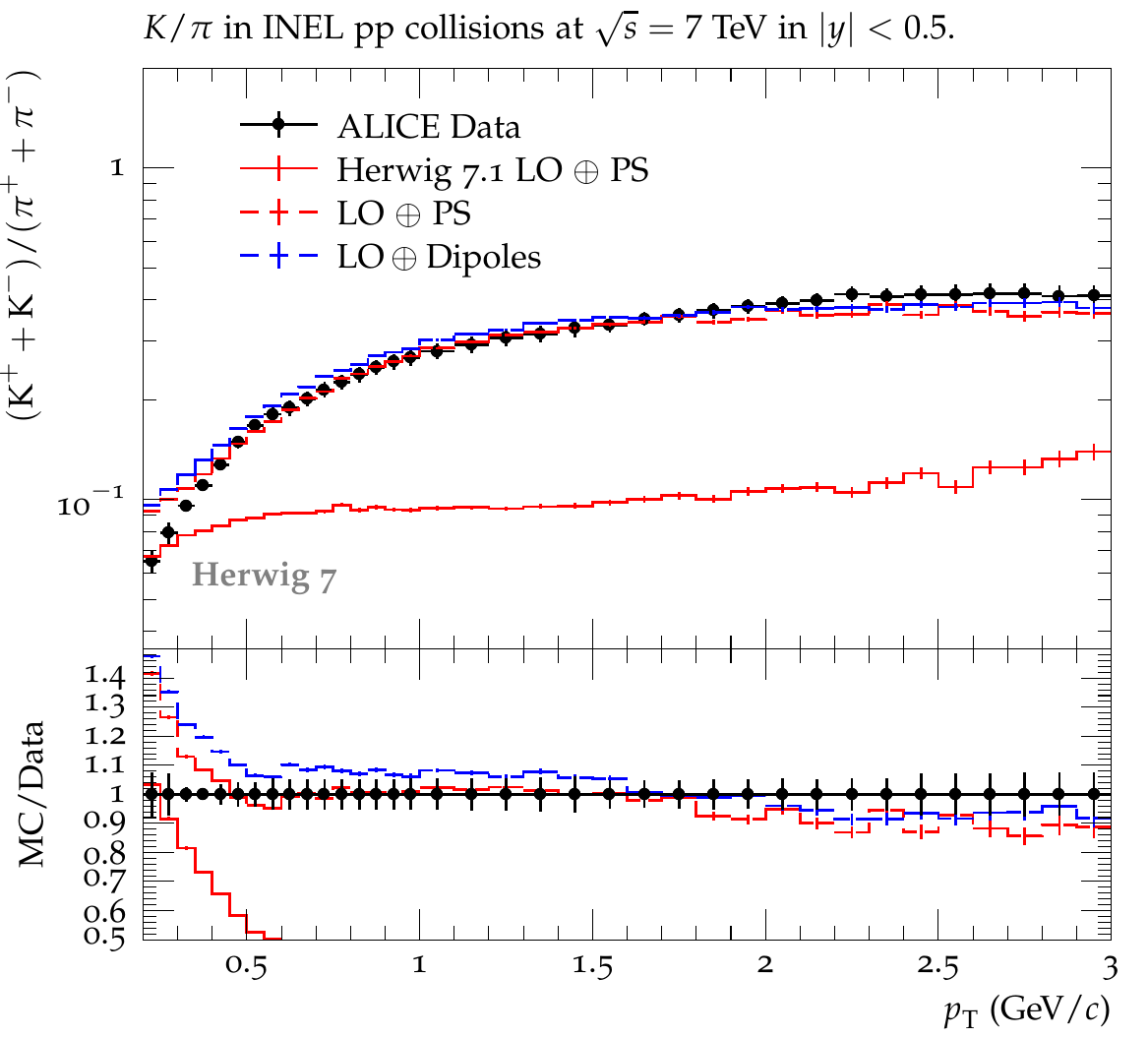}
\caption{The K to $\pi$ ratio in inelastic events in comparison with
ALICE data\cite{Adam:2015qaa}.
\label{fig:strangeness}}
\end{figure}

While the plain Colour Reconnection model~\cite{Gieseke:CRmodel} is an integral part of
the description of general properties of Minimum Bias (MB) data,
the description of flavour specific observables remained difficult. With Herwig 7.1.5 we
introduced a new Colour Reconnection model that reconnects clusters based on geometrical
properties. We also allow multiple mesonic clusters to form a \textit{baryonic}
type cluster if certain requirements are met. 
This gives an important lever on the baryon to meson ratio and proved to 
be a good starting point for the description of flavour observables. Additionally we allow
non-perturbative $g\rightarrow s\bar{s}$ splitting for an additional source of strangeness. 
With the new model, the whole range of MB data can be described with similarly good quality and the description of hadronic flavour observables improves significantly.
An example of the strangeness production is shown in
Fig.~\ref{fig:strangeness}, where we see a greatly improved description
of ALICE data with either of the shower models.
For more details on the implementation and the details of the model, we refer to~\cite{Gieseke:2017clv}.

\section{BSM Physics}

We have made significant improvements to the handling of models in the Universal FeynRules Output~(\textsf{UFO})
format. Previously we could only handle vertices that had the perturbative form of the interaction, for example
$(p_1-p_2)^\mu$ for vector-scalar-scalar interactions, where $p_{1,2}$ are the four momenta of the scalar particles.

We now make use of the \textsf{sympy} package~\cite{10.7717/peerj-cs.103} to allow us to write code capable of evaluating the HELAS building blocks
for arbitrary Lorentz structures. This allows \textsf{Herwig} to be used to simulate a much wider class of BSM models with,
for example, spin $\frac32$ particles, colour flows involving $\varepsilon$ tensors and sextet particles, and many
four-point interactions now supported. Splitting functions for the production of electromagnetic radiation are now also created by default for BSM particles.

\section{Modifications to multi parton interactions}
\label{sec:mpichanges}

\begin{figure}[t]
\includegraphics[width=0.48\textwidth]{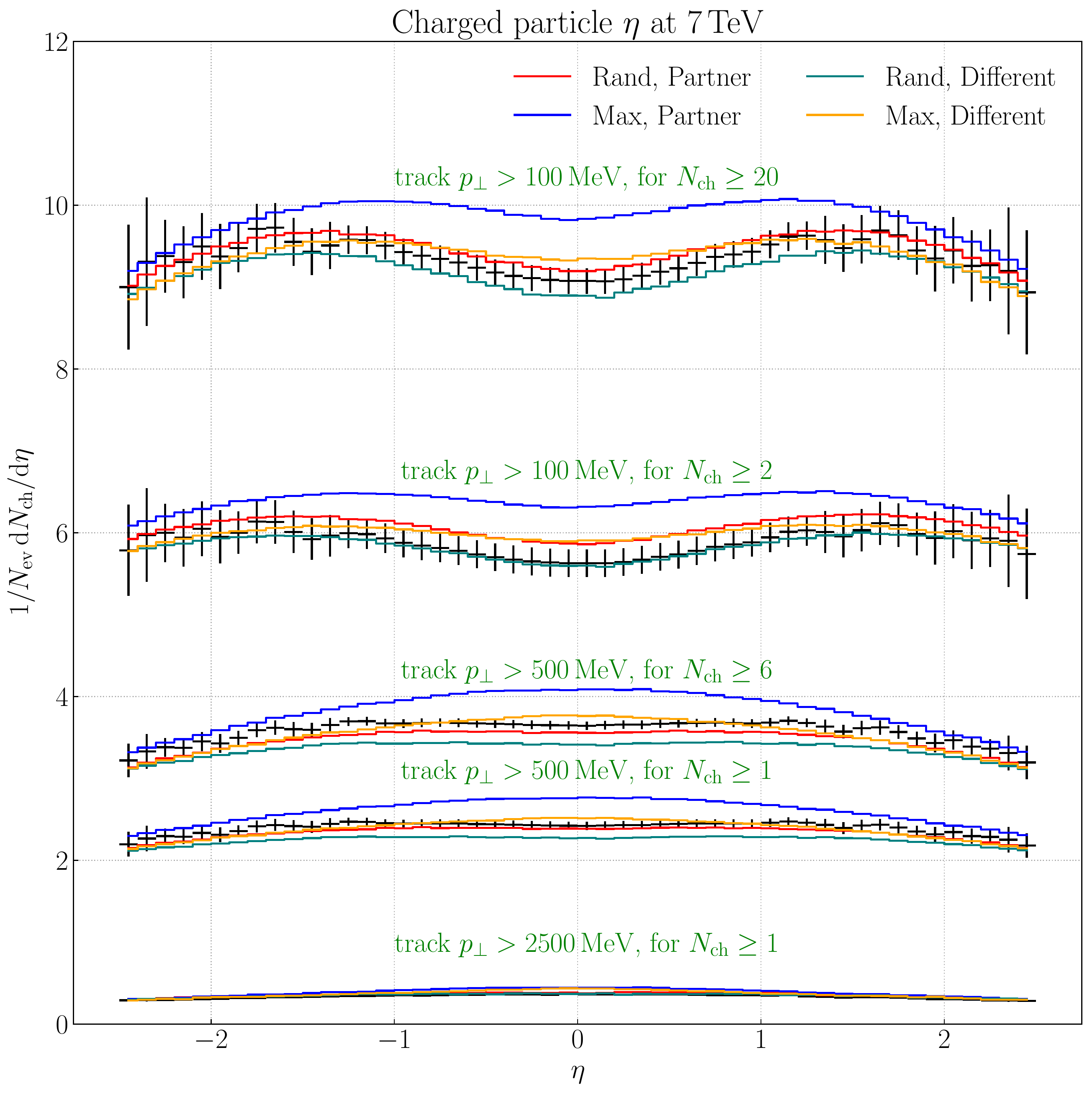}
\caption{The charged-particle multiplicity is plotted against the rapidity for multiple cuts (green labels) on the hardest track transverse momentum and number of charged particles. Data is taken from \cite{Aad:2010ac}. This observable is sensitive to the choices that are employed as the starting conditions of the parton shower process. The four choices are described in Sec. \ref{sec:mpichanges}. While height differences are easily modified in the tuning process, shape differences prefer choices with a random colour partner for gluons in the hard process.  \label{fig:partnerfinder}}
\end{figure}

While the space-time picture \cite{Bellm:2019wrh} is not included in the release various changes have been
 made to the handling of multiple parton interactions. The changes made are described in detail in
\cite{Bellm:2019icn}. Here, we give a summary of the findings and the implications for the newly released version.

The kinematics of the soft model have been modified to use the algorithm described in 
\cite{Jadach:1975woe}\footnote{The previous version of Herwig made use of algorithms described in  \cite{Baker:1976cv}. }, 
resulting in a disappearance of the unphysical correlation found in  \cite{Azarkin:2018cmr}.
Related to the kinematics of the soft ladders is the distribution that is used to generate the transverse momenta. 
Here, we allow switching between different schemes and we found that it is beneficial to produce the hardest parton in the ladder according to the old
distribution used in \cite{Bahr:2008pv} and the rest of the partons flat below this maximal value.  

The variable $p_\perp^{\min}$ which splits the hard from the soft scatterings
was found to give a good description of data at high energies if a power law was used to parametrize the energy dependence. 
At small centre-of-mass energies ( $\lesssim 200$ GeV ), this power-law generated values for
 $p_\perp^{\min}$ for which the eikonal model could not be solved. A comprehensive tuning effort 
 showed that a power-law with an offset can be used to describe the data and solve the model at any sensible
 energy. 

A `dummy' matrix element is used to start the production of minimal bias events. In this version, the processes handled by the ME are restricted to extract valence quarks only. The amount of forced splittings in the backward evolution to the incoming beams is therefore strongly reduced. 

We have replaced the cross-section reweighter, which was previously used,
and modified the matrix element used in minimum bias runs to reweight the cross section, such that the eikonalized cross sections are produced. 
This has the advantage of generating unit weights at the production level. 

Another change that is more on the technical side is the introduction of the parameter that controls the ratio of the diffractive cross-section as part of the inelastic cross-section,  named \texttt{DiffractionRatio}. It was previously a combination of the
\texttt{CSNorm} parameter and the construction of the matrix element weight. The new parameter allows a more controlled and physically motivated tuning. 

It was found that changed starting conditions for the showering of the gluons, in particular the recoil partner and scale choice, are beneficial for the description of charged multiplicities over rapidity. The default choice is the same as used previously in the showering of NLO matched samples and external LHE files. 
In Fig.~\ref{fig:partnerfinder} we illustrate the effect for the choices that choose the evolution partner randomly (\texttt{Rand}) or according to the maximal angle (\texttt{Max}) and allow the shower starting scale choice to be chosen according to the partner (\texttt{Partner}) or differently (\texttt{Different}).
 
 The combination of all the changes described here required a retuning of the MPI model. 
 Details are outlined in \cite{Bellm:2019icn}.

\section{Other Changes}

Besides the major physics improvements highlighted in the previous sections,
we have also made a number of smaller changes to the code and build system
which we will summarize below. Please refer to the online documentation for a
fully detailed description or contact the authors.

\subsection{SaS Parton Distribution Functions}

As version 6 of the \textsf{LHAPDF}~\cite{Buckley:2014ana} package does not contain any parton distributions
for the partons inside resolved photons the {\tt FORTRAN} code and an interface to the
Schuler-Sj\"ostrand~\cite{Schuler:1996fc} parton distribution functions for the photon have been
included to allow the simulation of resolved photon processes.

\subsection{FxFx}

The FxFx merging module was introduced in~\cite{Bellm:2015jjp} to
provide support of the NLO multi-jet merging method of
\cite{Frederix:2012ps}, via Les Houches-accord event (LHE) files generated
by \texttt{MadGraph~5/aMC@NLO}~\cite{Alwall:2014hca}.

In \Hw\ 7.2 this functionality is available by default, being
compiled with the main part of the code. The framework also
provides an interface for merging of tree-level events generated
either by \texttt{MadGraph~5/aMC@NLO} or \texttt{AlpGen} via the MLM
technique~\cite{Mangano:2002ea, Mangano:2004lu}, replacing all
the functionality that first appeared in~\cite{Arnold:2012fq}. The relevant input files for the FxFx merging and tree-level merging
are now \texttt{LHE-FxFx.in} and \texttt{LHE-MGMerging.in}
respectively. We emphasize that it is essential to include the MC@NLO
matching settings for \texttt{MadGraph~5/aMC@NLO} when performing the
FxFx merging, as given in \texttt{LHE-MCatNLO.in}. These settings
should not be included when merging tree-level events. The tree-level
merging functionality via \texttt{MadGraph~5/aMC@NLO} events uses the
event tags in the appropriately-generated LHE files and requires the option
\texttt{MergeMode} to be set to \texttt{TreeMG5}, as is done by
default in \texttt{LHE-MGMerging.in}. To enable merging with events generated via \texttt{AlpGen}, 
\texttt{MergeMode} should be switched to \texttt{Tree}.

We note that the FxFx functionality has been tested thoroughly only for $\mathrm{W+jets}$
and $\mathrm{Z+jets}$ events in~\cite{Frederix:2015eii}, where it was
compared against LHC data at 7 and 8 TeV. We also note that no tuning
was performed in \Hw\ using events generated via this
interface.

\subsection{Default PDF}

  The default parton distribution function has been changed from that of MMHT 2014~\cite{Harland-Lang:2014zoa} to CT14~\cite{Dulat:2015mca}.

\subsection{Minor improvements and bug fixes}

A number of minor changes and bug fixes are worth noting, in particular, there
have been new options for the physics simulation besides the ones described in
the previous text:
\begin{itemize}
\item major updates in the {\tt Tests} directory to improve both the generation of input files and
  add new \Rivet\ analyses.
\item a number of changes have been made to ensure that the \Hw\ code compiles with
  the \textsf{Intel} and \textsf{Clang} compilers. A number of changes have also been made to
  ensure compilation with recent \textsf{gcc} compilers, including \textsf{gcc9}.
\item The deprecated \textsf{UA5} soft underlying event model has been removed.
\item The input files for a number of old tunes have been removed.
\item The cut-off for photon radiation from leptons has been reduced to  $10^{-6}$ GeV.
\item Support for fixed target collisions has been included, together with an example input file.
\item The analytic calculation of the partial width for $V\to SS$ decays has been corrected.
\item The setting of masses in \textsf{UFO} models where one parameter sets the masses of many particles has been fixed.
\item An effective vertex for the processes $h^0\to Z^0\gamma$ has been added so the $Z^0$ mass is correctly generated in this decay.
\item Fix to the \textsf{MEvv2vs} class so that more than one four-point vertex is allowed.
\item A missing $t$-channel diagram has geen added to the \textsf{MEfv2fs} class.
\item Changes to avoid $0/0$ have been made in the \textsf{VVVDecayer} class.
\item An option to use the internal Standard Model Higgs boson vertices for UFO models which do not implement the full Higgs sector has been added.
\item Several bugs in the presence of spacelike off-shell incoming
legs have been fixed in \ThePEG's \texttt{StandardXComb} and
\Hw's \texttt{Tree2toNPhasespace} classes.
\item The option of an asymetric splitting of the colour flows for the $g\to gg$ branching
  in the dipole shower has been added.
\item Additional kernels are implemented for the $\tilde{q}$ shower to incorporate the Catani-Marchesini-Webber~(CMW) scheme as part of a linear scheme. By default, the scheme is absorbed in a change of the nominal value of the strong coupling  $\alpha_S(M_Z)$. A similar scheme has been available for the dipole shower since Herwig 7.1.
 \item The dipole shower has been tuned using the method described in \cite{Bellm:2019owc}.
\end{itemize}
Technical issues which have been addressed include:
\begin{itemize}
\item The old {\tt ClassTraits} mechanism used by \ThePEG\ has been replaced by the new {\tt DescribeClass}
  mechanism consistently in all the \Hw\ code.
\item Changes to the templates for dimension-full quantities to improve the maintainability of this code. Regrettably
  this is incompatible with \textsf{gcc 4.8} and therefore \textsf{gcc 4.9} is now the oldest supported version of \textsf{gcc}.
    \item A number of changes have been made to ensure the bootstrap script works with \textsf{python3}, however a number of
      our dependencies do not yet support \textsf{python3} and therefore the code still uses \textsf{python2}.
    \item The generation of trial values of the scale and light-cone momentum fraction in the angular-ordered parton shower has been restructured to improve performance.
    \item The calculation of the cross section in Matchbox processes has been restructured to reduce calls to the parton distribution functions, and hence improve performance.
    \item Changes have been made to improve the detection of recent \textsf{boost} versions at compile time.
    \item A number of changes have been made to our test suite to include more \textsf{Rivet} analyses and improve
    the output of the results.
\end{itemize}

\subsection{Build and external dependencies}

Since version 7.1, \Hw\ has enforced the use of a C++11 compliant
compiler, and C++11 syntax and standard library functionality is used widely
within the code. The \texttt{herwig-bootstrap} script is able to provide such
a compiler along with a full \Hw\ plus dependencies build.
\texttt{herwig-bootstrap} will also enforce the newest versions of external
amplitude providers; specifically we now use:
\begin{itemize}
\item \OL\ \cite{Cascioli:2011va} versions $\ge$ 2.0.0 with the
  \Collier\ library \cite{Denner:2016kdg} for tensor reduction (should older
  versions of \OL\ be required, the input files require the additional option
  \texttt{set OpenLoops:UseCollier Off}), and
\item \GoSam\ versions $\ge$ 2.0.4 to pick up the correct normalization for
  loop induced processes outside of specialized setups.
\end{itemize}
A number of changes have also been implemented to reduce run-time load for
allocating and de-allocating various containers, and to reduce overall memory
consumption.

\subsection{Licensing}

While older versions were licensed under the GNU
General Public License GPL version 2, since version~7.1, \Hw\ has been
distributed with the GPL version~3. The MCnet guidelines for the distribution
and usage of event generator software in an academic setting apply as before,
and both the legally binding GPL license and the MCnet guidelines are
distributed with the code.

\section{Example Results}

\Hw~7.2 has been thoroughly validated against a wide range of existing
data, as implemented in the Rivet and FastJet
frameworks\cite{Buckley:2010ar,Cacciari:2011ma}. Parameter tuning has
been performed using Professor\cite{Buckley:2009bj}.

\begin{figure}[t]
\includegraphics[width=0.48\textwidth]{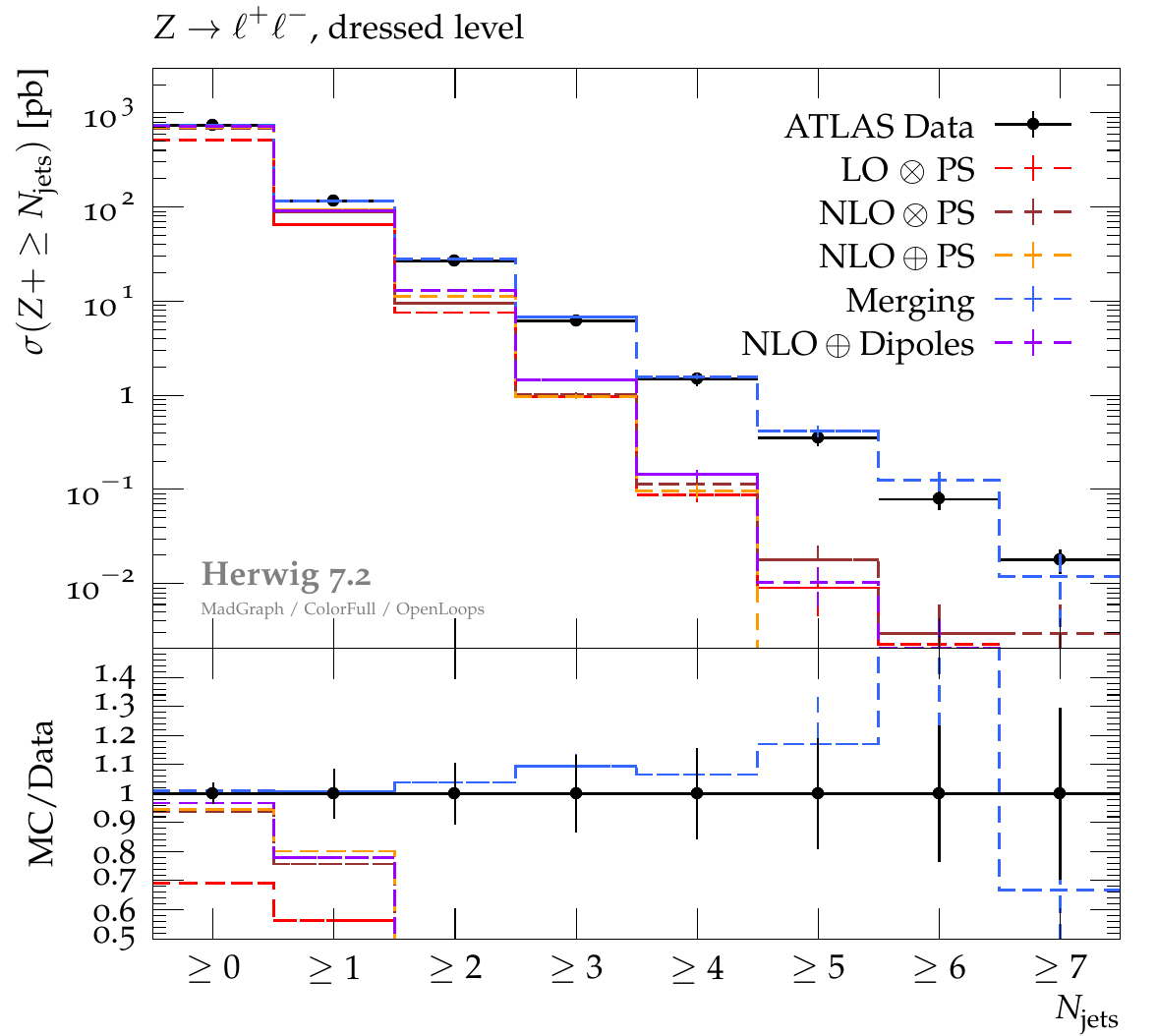}
\includegraphics[width=0.48\textwidth]{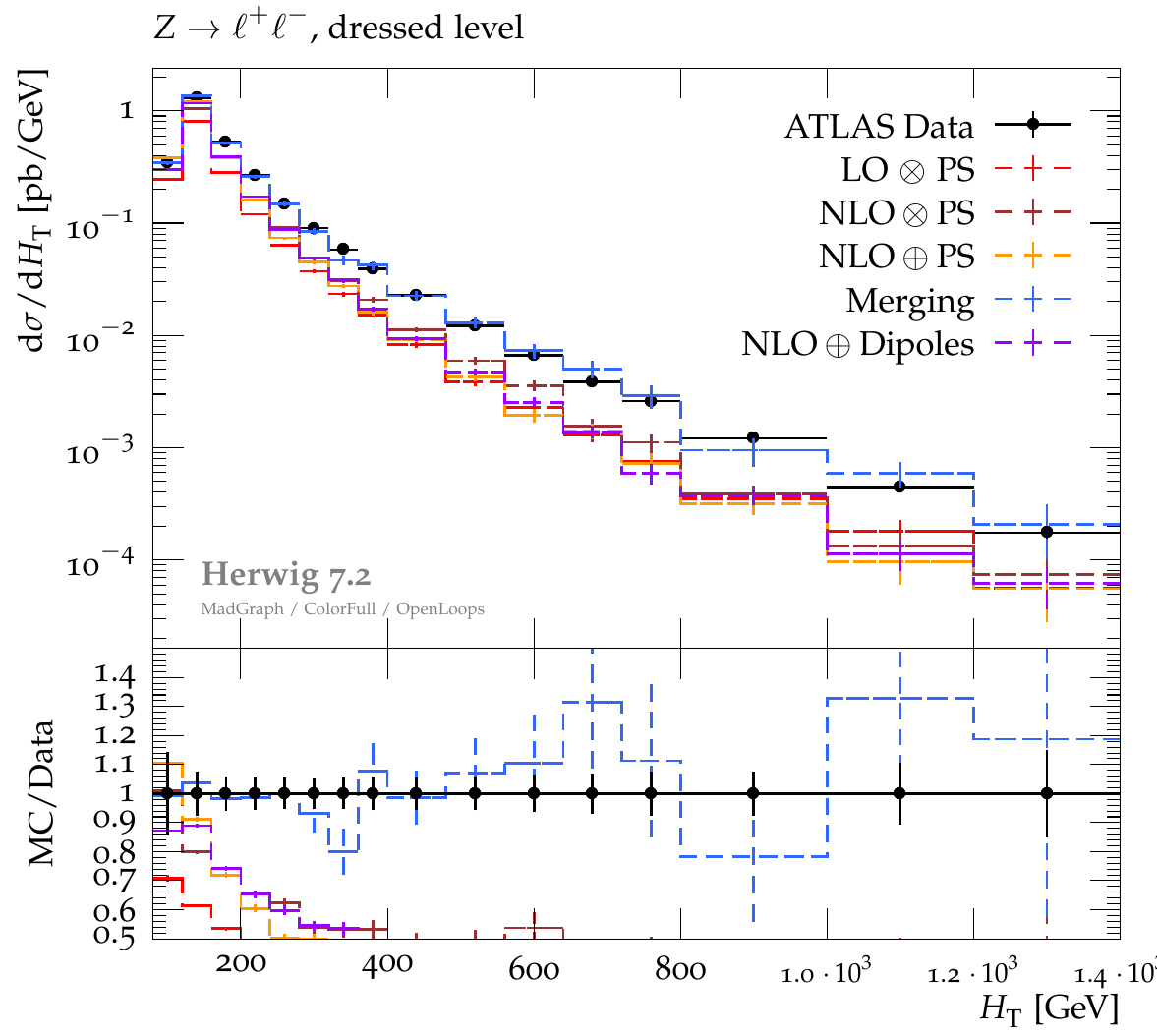}
\caption{The cross section for Z production in association with
$N_{\textrm{jets}}$ jets (upper) or differentially with respect to the
total scalar sum of final state transverse momenta, $H_{\textrm{T}}$,
(lower) in comparison with ATLAS data\cite{Aaboud:2017hbk}.
\label{fig:multijet}}
\end{figure}

Here, we illustrate some examples of the fact that we can simulate LHC
events with any combination of LO or NLO matrix elements, matched with
the angular-ordered or dipole showers using either additive
(MC@NLO-like) or multiplicative (POWHEG-like) methods, as well as
multi-jet merging, for Z boson production. In Fig.~\ref{fig:multijet},
we show the results in comparison with ATLAS
data\cite{Aaboud:2017hbk}.
The upper plot shows that, as would be hoped, merging with multi-jet
matrix elements enables a good description of the data over a wide range
of jet mupliticities. The lower plot shows that even for more inclusive
quantities, such as the total scalar transverse momentum, the multijet
effects are important.

A wide range of further plots can be found at
\texttt{\href{https://herwig.hepforge.org/plots/herwig7.2}{https://herwig.hepforge.org/plots/herwig7.2}}.

\section{Summary and Outlook}

We have described a new release, version 7.2, of the \Hw\ event
generator. This new release contains a number of improvements to both
perturbative and non-perturbative simulation of collider physics and will form
the basis of further improvements to both physics and technical aspects.

\section*{Acknowledgments}

We are indebted to Leif L\"onnblad for his authorship of \ThePEG, on which
\Hw\ is built, and his close collaboration, and also to the authors of
\Rivet\ and \textsf{Professor}. We are also grateful to Malin Sj\"odahl for
providing the \textsf{ColorFull} library for distribution along with \Hw.

This work was supported in part by the European Union as part of the 
H2020 Marie Sk\l odowska-Curie Initial Training Networks MCnetITN3
(grant agreement no. 722104) and the UK Science and Technology Facilities Council (grant numbers
ST/P000800/1, ST/P001246/1). 
JB and CR are part of the MorePheno project that is funded by the European Research 
Council (ERC) under the European Union's Horizon 2020 
research and innovation programme, grant agreement No 668679. 
AP acknowledges support by the ERC grant ERC-STG-2015-677323. 
GB thanks the UK Science and
Technology Facilities Council for the award of a studentship.
A.S. acknowledges support from the National Science
Centre, Poland Grant No. 2016/23/D/ST2/02605.
This work has been supported by the BMBF under grant number 05H18VKCC1.

\bibliography{Herwig}
\end{document}